\documentclass[10pt,superscriptaddress,twocolumn,amsmath,amssymb,aps,pra,showpacs]{revtex4-1}
\usepackage{mathrsfs}
\usepackage{graphicx}
\usepackage{dcolumn}
\usepackage{bm}
\usepackage[title,titletoc,header]{appendix}

\usepackage{amssymb}
\usepackage{amsmath}
\usepackage{mathrsfs}
\usepackage{amsmath}
\usepackage{subfigure}
\usepackage{float}
\usepackage[title,titletoc,header]{appendix}
\usepackage{graphicx}

\newcommand{\be}{\begin{equation}}
\newcommand{\ee}{\end{equation}}
\newcommand{\bea}{\begin{eqnarray}}
\newcommand{\eea}{\end{eqnarray}}

\begin{document}
\title{Infinite bound states and hydrogen atom-like energy spectrum induced by a flat band }
\author{Yi-Cai Zhang}
\affiliation{School of Physics and Materials Science, Guangzhou University, Guangzhou 510006, China}

\author{Guo-Bao Zhu}
\affiliation{School of  Physics and Electronic Engineering, Heze University, Heze 274015, China}


\date{\today}
\begin{abstract}
In this work, we investigate the bound state problem in  one dimensional spin-1 Dirac Hamiltonian  with a flat band. It is found that,   the flat band has significant effects on the bound states. For example,  for Dirac delta potential $g\delta(x)$, there exists one bound state for both positive and negative potential strength $g$.  Furthermore, when the potential  is weak, the bound state energy is proportional to the potential strength $g$.
For square well potential, the flat band results in the existence of infinite bound states for arbitrarily weak potential. In addition, when the bound state energy is very near the flat band, the energy displays hydrogen atom-like spectrum, i.e., the bound state energies are inversely proportional to the square of natural number $n$ (e.g., $E_n\propto 1/n^2, n=1,2,3,...$). Most of  the above nontrivial behaviors can be attributed to the infinitely large density of states of flat band and its ensuing $1/z$ singularity of Green function. The combination of a short-ranged potential and flat band provides a new possibility to get infinite number of bound states and hydrogen atom-like energy spectrum. In addition, our findings would provide some useful insights in the understanding of many-body physics of flat band.

\end{abstract}

\maketitle
\section{Introduction}

In the past decades, the physics induced by flat bands have attracted great interests \cite{Sutherland1986,Vidal1998,Bergman2008,Bercioux2009,Moessner2011,Wang2011,Raoux2014,Peotta2015,Illes2016,Hazra2019,Leykam2018,Cao2018,Chen2019,Xu2020,Wuyurong2021}.
In contrast with ordinary dispersion bands, a lot of  novel phenomena, for example, the ferro-magnetism transition \cite{Mielke1999}, localization \cite{Leykam2017}, super-Klein tunneling \cite{Shen2010,Urban2011,Fang2016,Ocampo2017,Illes2017}, quantum Hall-like states \cite{BERGHOLTZ,Yang2012}, zitterbewegung\cite{Ghosh}, preformed pairs \cite{Tovmasyan2018}, strange metal \cite{Volovik2019}, high-$T_c$ superconductivity/superfluidity \cite{Kopnin2011,Julku2020,Hu2019,Xie2020}, ect., would appear.
Flat bands have some prominent features worthy of notice. First of all, the states of flat band have localized properties. For example, the states  only occupy a finite number of unit cells in lattice model \cite{Mukherjee}. Secondly, the flat band is very sensitive to small perturbations, e.g., the magnetism transitions driven by a interaction  \cite{Wu2008,Zhang2010}. Thirdly, the infinitely large density of states can give rise to  a linear dependence of interaction strength of superconductor/superfluid order parameters \cite{Iglovikov2014,Julku2016,Liang2017,Iskin2019,Wu2021}.

%
%

It is well known that the behaviors of density of states near the threshold of the continuous energy spectrum play crucial roles in the formations of bound states \cite{Economou}. For example, in three-dimensional case, the density of states near the threshold ($E_{th}=0$)  follows the law $\rho_0(E)\propto \sqrt{E}\rightarrow 0$ as $E\rightarrow 0$, then only when the potential strength is sufficiently large [for a square well potential with width $a>0$,  the depth $|V|>\pi^2\hbar^2/(8 m a^2)$] \cite{Landau}, a bound state can exist. While for a two-dimensional counterpart, due to a finite density of states near band edge, there exist bound states no matter how weak  an attractive potential is. For a one-dimensional case, owing to the divergence of density of state near band edge [$\rho_0(E)\propto 1/\sqrt{E}\rightarrow\infty $ as $E$ goes to zero], the bound states also exist for arbitrarily weak attractive potential. In addition, in the presence of weak potential well,  the shallow bound state energy is proportional to the square of potential strength. Due to the singular density of density of states of flat band, it is expected that the existence conditions of bound states would be affected significantly.


The bound state problem in two-dimensional flat band model with a short-ranged potential well and a lang-ranged Coulomb potential have been investigated by  Gorbar et al. \cite{Gorbar2019}. It is found that for arbitrary weak potential, there exists bound states which split from the flat band. In the presence of Coulomb potential, the flat band could not survive.
Furthermore, Pottelberge found that the flat band  gradually evolves into a continuous band  (uncountable set) with the increase of Coulomb potential strength \cite{Pottelberge2020}.

In this work, we investigate the bound states in a one-dimensional spin-1 Dirac-type Hamiltonian.
It is found that for an arbitrarily weak delta potential, similarly as that in two-dimensional case \cite{Gorbar2019}, there exists a bound state which is generated from the flat band. Differently from the ordinary one-dimensional case,  the bound state energy is linearly dependent on the potential strength as the strength goes to zero.
 It is believed that the existence of a hydrogen atom-like  energy spectrum $E_n\propto 1/n^2$ \textbf{\cite{Landau}} usually requires a long-ranged Coulomb potential $1/r$ .
Surprisingly for the flat band system, even a short-ranged square well potential can result in an infinite number of bound states (countable set).  The existence of infinite bound states originates from an arbitrarily strong effective potential induced by the flat band. In addition, when the bound state energy is much smaller than the typical energy scale of the system, it displays a hydrogen atom-like energy spectrum, i.e., $E_n\propto 1/n^2$.

The work is organized as follows. In Sec.\textbf{II}, the three energy bands, free particle Green function and density of states are investigated.  Next, we solve the bound state problems for several typical short-ranged potentials in Sec.\textbf{III}.
 At the end, a summary is given in Sec.\textbf{IV}.
\begin{figure}
\begin{center}
\includegraphics[width=1.0\columnwidth]{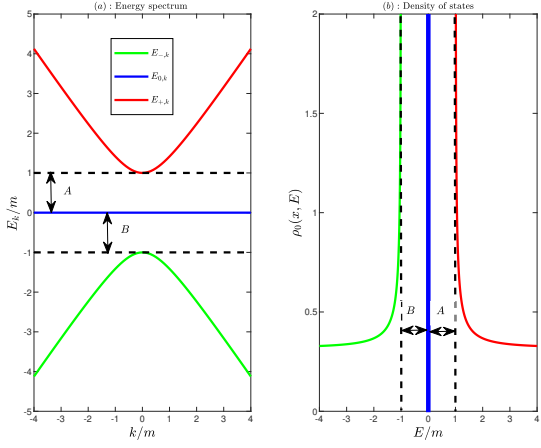}
\end{center}
\caption{ (a): The energy spectrum of free particle Hamiltonian. (b): Density of states of the spin-1 Dirac model. The flat band gives rise to an infinitely large density of states at zero energy (thick blue lines). Near the thresholds  of upper and lower bands $E_{th}=\pm m$, the density of states has a similar divergence as that in the ordinary one-dimensional case, i.e., $\rho_0(x,E)\propto 1/\sqrt{|E-E_{th}|}\rightarrow \infty$ as $E\rightarrow E_{th}$.  Any possible bound states only exist in the regions \textbf{A} and \textbf{B}. }
\label{schematic}
\end{figure}
%

\section{The model Hamiltonian  with a flat band}

In this work, we consider a three component spin-1 Dirac-type Hamiltonian \cite{Ocampo2017} in one dimension, i.e.,
 \begin{align}
&H=H_0+V_p(x)\notag\\
&H_0=-iv_F\hbar S_x\partial_x+m S_z,
    	\label{hamiltonian}
\end{align}
where $V_p(x)$ is potential energy, $H_0$ is the free-particle Hamiltonian, $v_F>0$ is Fermi velocity, and $m>0$ is energy gap parameter. $S_x$ and $S_z$ are spin operators for spin-1 particles \cite{Zhang2013,Chenxiaomei2019}, i.e.,
\begin{align}
&S_x=\left[\begin{array}{ccc}
0 &\frac{1}{\sqrt{2}}  & 0\\
\frac{1}{\sqrt{2}}& 0 &\frac{1}{\sqrt{2}}\\
0 &\frac{1}{\sqrt{2}} & 0
  \end{array}\right];&S_z=\left[\begin{array}{ccc}
1 &0  & 0\\
0&0& 0\\
0 &0 & -1
  \end{array}\right],
\end{align}
in usual basis $|i\rangle$ with $i=1,2,3$. In the whole manuscript, we use the units of $v_F=\hbar=1$. The above Hamiltonian may be realized in photonic systems \cite{Huang2011,Chan2012}.
 When $V_p(x)=0$, for a given momentum $k$, the free particle Hamiltonian $H_0$ has three eigenstates  and the
 eigenenergies, i.e.,
\begin{align}
&\langle x |-,k\rangle=\psi_{-,k}(x)=\frac{1}{2\sqrt{k^2+m^2}}\left(\begin{array}{ccc}
\sqrt{k^2+m^2}-m\\
 -\sqrt{2}k\\
\sqrt{k^2+m^2}+m
  \end{array}\right)e^{ikx},\notag\\
  &E_{-,k}=-\sqrt{k^2+m^2};\notag\\
  &\langle x |0,k\rangle=\psi_{0,k}(x)=\frac{1}{\sqrt{2(k^2+m^2)}}\left(\begin{array}{ccc}
-k\\
 \sqrt{2}m\\
k
  \end{array}\right)e^{ikx},\notag\\
  &E_{0,k}=0;\notag\\
  &\langle x |+,k\rangle=\psi_{+,k}(x)=\frac{1}{2\sqrt{k^2+m^2}}\left(\begin{array}{ccc}
\sqrt{k^2+m^2}+m\\
 \sqrt{2}k\\
\sqrt{k^2+m^2}-m
  \end{array}\right)e^{ikx},\notag\\
  &E_{+,k}=\sqrt{k^2+m^2}.
\end{align}
$|-,0,+; k\rangle$  denote the states for the lower, middle (flat band) and upper bands, respectively and  $E_{-,k},E_{0,k}, E_{+,k}$ represent the three corresponding energy bands.
It is found that a flat band with zero energy ($E_{0,k}=0$) appears in between upper and lower bands (see Fig.1). The possible bound states (if any) can only exist in the gaps among the three bands, i.e., $0<E<m$ and $-m<E<0$ (the regions \textbf{A} and \textbf{B} in the Fig.1).

In order to explore the properties of the flat band, we calculate the free particle Green function $G_0$ (corresponding to $H_0$) and density of states $\rho_0(x,E)$ \cite{Economou}.
In momentum space, the free particle Green function can be expressed as
\begin{align}
&G_0(k,k',z)\equiv\langle k|\frac{1}{z-H_0}|k'\rangle\notag\\
&=\left[\begin{array}{ccc}
G_{0,11}(k,k',z) &G_{0,12}(k,k',z) & G_{0,13}(k,k',z)\\
G_{0,21}(k,k',z)&G_{0,22}(k,k',z) &G_{0,23}(k,k',z)\\
G_{0,31}(k,k',z) &G_{0,32}(k,k',z) & G_{0,33}(k,k',z)
  \end{array}\right],\notag\\
  &=\frac{\delta(k-k')}{2(m^2z+k^2z-z^3)}\notag\\
  &\times\left[\begin{array}{ccc}
k^2-2z(m+z) &-\sqrt{2}k(m+z) &-k^2\\
-\sqrt{2}k(m+z)&2(m^2-z^2) &\sqrt{2}k(m-z)\\
-k^2 &\sqrt{2}k(m-z) & k^2+2z(m-z)
  \end{array}\right].
\end{align}

In coordinate space, the matrix elements of the free particle Green function are \cite{Zhang2014}
\begin{align}\label{H0}
G_{0,ij}(x,x',z)=\frac{1}{2\pi}\int_{-\infty}^{\infty} dkdk'e^{i(kx-k'x')}G_{0,ij}(k,k',z),
\end{align}
or explicitly written down
\begin{align}\label{6}
&G_{0,11}(x,x',z)=\frac{-(m+z)}{2\sqrt{m^2-z^2}}e^{-\sqrt{m^2-z^2}|x-x'|}\notag\\
&-\frac{\sqrt{m^2-z^2}}{4z}e^{-\sqrt{m^2-z^2}|x-x'|}+\frac{\delta(x-x')}{2z},\notag\\
&G_{0,12}(x,x',z)=\frac{-i(m+z)sign(x-x')}{2\sqrt{2}z}e^{-\sqrt{m^2-z^2}|x-x'|},\notag\\
&G_{0,13}(x,x',z)=\frac{\sqrt{m^2-z^2}}{4z}e^{-\sqrt{m^2-z^2}|x-x'|}-\frac{\delta(x-x')}{2z},\notag\\
&G_{0,21}(x,x',z)=G_{0,12}(x,x',z)\notag,\\
&G_{0,22}(x,x',z)=\frac{\sqrt{m^2-z^2}}{2z}e^{-\sqrt{m^2-z^2}|x-x'|},\notag\\
&G_{0,23}(x,x',z)=\frac{i(m-z)sign(x-x')}{2\sqrt{2}z}e^{-\sqrt{m^2-z^2}|x-x'|},\notag\\
&G_{0,31}(x,x',z)=G_{0,13}(x,x',z),\notag\\
&G_{0,32}(x,x',z)=G_{0,23}(x,x',z),\notag\\
&G_{0,33}(x,x',z)=\frac{(m-z)}{2\sqrt{m^2-z^2}}e^{-\sqrt{m^2-z^2}|x-x'|}\notag\\
&-\frac{\sqrt{m^2-z^2}}{4z}e^{-\sqrt{m^2-z^2}|x-x'|}+\frac{\delta(x-x')}{2z},
\end{align}
where the sign function $sign(x)=1$ for $x>0$, $sign(x)=-1$ for $x<0$ and
 $\delta(x)$ is Dirac delta function.
The Dirac delta term $\delta(x-x')/(2z)$ in the Green function arises from the flat band, which reflects the localized properties of states in flat band \cite{Mukherjee}.

The density of states at energy $E$ (per unit length)  is given by the imaginary part of Green function \cite{Economou}, i.e.,
 \begin{align}
    	&\rho_0(x,E)=-lim_{x'\rightarrow x}\frac{Im\{Tr[G_0(x,x',z=E+i0_+)]\}}{\pi}\notag\\
        &=-\frac{Im\{\sum_{i=1,2,3}G_{0,ii}(x,x,z=E+i0_+)\}}{\pi}\notag\\
        &=-\frac{Im[\frac{\delta(0)}{E+i0_+}+\frac{-(E+i0_+)}{\sqrt{m^2-(E+i0_+)^2}}]}{\pi}\notag\\
        &=\frac{\theta(E-m)|E|}{\pi \sqrt{E^2-m^2}}+\delta(0)\delta(E)+\frac{\theta(-E-m)|E|}{\pi \sqrt{E^2-m^2}},
    	\label{7}
\end{align}
where $\theta(x)=1$ if $x>0$, otherwise $\theta(x)=0$ and $\delta(0)=lim_{x'\rightarrow x}\delta(x-x')\rightarrow\infty$.

 It shows that the flat band of zero energy gives rise to an infinitely large  density of states [$\delta(0)$] \cite{Gorbar2019}, which reflects the fact that  there are infinite states of flat band in the continuous model. Furthermore, the $\delta(E)$ singularity in density of states $\rho_0(x,E)$ causes a $1/z$ singularity in the Green function [see Eq.(\ref{6}) and \textbf{Appendix A}].   In addition, near the thresholds of continuous spectrum ($E_{th}=\pm m$), the density of states shows a similar divergence as  that in ordinary one-dimensional case, e.g., $\rho_0(x,E)\propto 1/\sqrt{|E-E_{th}|}\rightarrow \infty $ as $E\rightarrow E_{th}$ [see Fig.1].

In the following,  we would see that  such an infinitely large  density of states and $1/z$ singularity of the Green function have important effects on bound states (see \textbf{Appendix B}). We can naively think that,  due to the non-dispersion of flat band, an arbitrarily weak potential may pull some zero-energy states out the flat band, then  these states form bound states. It is shown that, indeed for some kinds of square well potentials, there exist infinite bound states for arbitrarily weak potential (see the next section).

\section{bound states}
In the following manuscript, we assume the potential energy $V_p$ has following diagonal form in usual basis $|i=1,2,3\rangle$, namely,
\begin{align}\label{V}
&V_p(x)=V_{11}(x)\bigotimes|1\rangle\langle1|+V_{22}(x)\bigotimes|2\rangle\langle2|\notag\\
&+V_{33}(x)\bigotimes|3\rangle\langle3|
=\left[\begin{array}{ccc}
V_{11}(x) &0  & 0\\
0&V_{22}(x)& 0\\
0 &0 & V_{33}(x)
  \end{array}\right].
\end{align}

\subsection{Delta potential}
First of all, we assume the potential energy has the form of the delta potential and satisfies  $V_{11}(x)=V_{33}(x)\equiv0$,  and $V_{22}(x)=g\delta(x)$ with potential strength $g$.
The Schr\"{o}dinger equation ($H\psi=E\psi$) can be written in terms of three component wave functions, i.e.,
\begin{align}\label{9}
&-i\partial_x\psi_2(x)/\sqrt{2}=[E-m]\psi_1(x),\notag\\
&-i\partial_x[\psi_1(x)+\psi_3(x)]/\sqrt{2}=[E-g\delta(x)]\psi_2(x),\notag\\
&-i\partial_x\psi_2(x)/\sqrt{2}=[E+m]\psi_3(x).
\end{align}

Furthering using Eq.(\ref{9}) to eliminate wave functions for 1-th and 3-th components, we get an effective Schr\"{o}dinger  equation (second order differential equation) for $\psi_2(x)$
\begin{align}\label{10}
\partial^{2}_x\psi_2(x)+\frac{[E-g\delta(x)](E^2-m^2)}{E}\psi_2(x)=0.
\end{align}

For bound states and $x\neq0$, the wave function $\psi_2(x)$  can be written as
\begin{align}\label{11}
&\psi_2(x)=Ce^{-\lambda x} & x>0,\notag\\
&\psi_2(x)=De^{\lambda x} & x<0,
\end{align}
where $\lambda=\sqrt{m^2-E^2}>0$, $C$ and $D$ are two coefficients for decayed part of wave function.
The continuity of  $\psi_2(x)$ at origin $x=0$ requires $C=D$. Integrating the both sides of Eq.(\ref{10}) near the origin $x=0$, it is shown that the derivative of $\psi_2(x)$ satisfies
\begin{align}\label{12}
&\psi'_2(0^+)-\psi'_2(0^-)=\frac{g(E^2-m^2)}{E}\psi_2(0),\notag\\
&\rightarrow-2\lambda C=\frac{g(E^2-m^2)}{E}C.
\end{align}
Based on  Eq.(\ref{12}), the bound state energy $E_B$ is obtained
\begin{align}\label{13}
&E_B\equiv E=\frac{mg}{\sqrt{4+g^2}}.
\end{align}

First of all, there always exists one bound state no matter the potential is repulsive ($g>0$) or attractive ($g<0$) [see Fig.(2)]. While for ordinary one dimensional case, only when the potential is attractive ($g<0$), there  is one bound state.  Secondly, when potential strength $g$ goes to zero, the bound state energy $E_B$ is linearly proportional to the strength, i.e., $E_B\propto g$ as  $g\rightarrow 0$ [see  Panel (b) of Fig.2]. Such a behavior is very different from the ordinary one-dimensional bound state energy, which is proportional to $g^2$. The linear dependence on the strength is also consistent with the fact  that the superfluid/superfluidor order parameter is linearly proportional to the two-body interaction strength in a flat band superfluid system \cite{Iglovikov2014,Julku2016,Julku2020}. Most of all these peculiar behaviors are due to  the infinitely large density of states of flat band (see \textbf{Appendix B}).

Thirdly, due to the restrictions of the thresholds of the upper and lower continuums, when the potential strength is very large ($g\rightarrow \infty$), the bound state energy $E_B$ approaches the thresholds of continuums $\pm m$ asymptotically. This is also very different from the usual one dimensional bound state energy which goes to infinity  ($\propto g^2\rightarrow\infty $).

\begin{figure}
\begin{center}
\includegraphics[width=1.0\columnwidth]{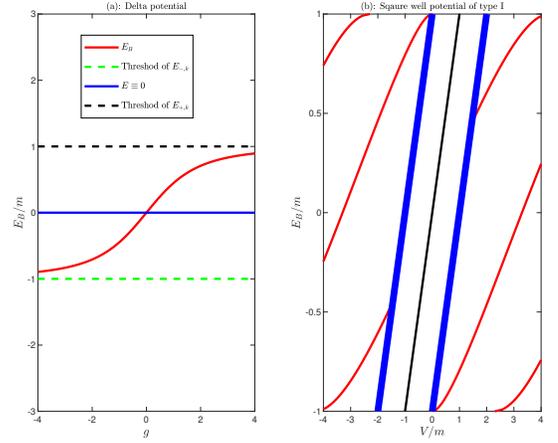}
\end{center}
\caption{ The bound state energies (the red solid lines).  Panel (a) is the bound state energy of delta potential. Panel (b) is the bound state energy of square well potential of type I (the red solid lines).  Near the thresholds of upper and lower continuums  $E_{th}=\pm m$ ($|E-m|/m\ll1$), the bound state energy can be approximated by the Eq.(\ref{22}). The (non-trivial) bound states only appear in the regions defined by $(E-V)^2-m^2>0$ [see the Eq.(\ref{18})].   The two thick blue solid lines are the boundaries of these regions  [$(E-V)^2-m^2=0$].  The black dashed solid line corresponds the trivial bound states, which is basically the original states of flat band with a shifted energy $V$ (see the main text). In panel (b), we take the width of potential well $a=1/m$. }
\label{schematic}
\end{figure}


\subsection{Square well potential of type I}
In this subsection, we assume the potential satisfies  $V_{11}(x)=V_{22}(x)=V_{33}(x)$  and
\begin{align}\label{14}
&V_{11}(x)=0 & x>a/2,\notag\\
&V_{11}(x)=V & -a/2<x<a/2,\notag\\
&V_{11}(x)=0 & x<-a/2,
\end{align}
where $a>0$ and $V$ are  the width  and depth (or height) of square well potential, respectively.
Now the Schr\"{o}dinger  equation is
\begin{align}\label{15}
&-i\partial_x\psi_2(x)/\sqrt{2}=[E-m-V_{11}(x)]\psi_1(x),\notag\\
&-i\partial_x[\psi_1(x)+\psi_3(x)]/\sqrt{2}=[E-V_{11}(x)]\psi_2(x),\notag\\
&-i\partial_x\psi_2(x)/\sqrt{2}=[E+m-V_{11}(x)]]\psi_3(x).
\end{align}
Similarly as above subsection, we get an effective second order differential equation
\begin{align}\label{16}
& \partial^{2}_x\psi_2(x)+\partial_x [ln(\frac{E-V_{11}(x)}{(E-V_{11}(x))^2-m^2})]\partial_x\psi_2(x)\notag\\
&+[(E-V_{11}(x))^2-m^2]\psi_2(x)=0.
\end{align}
In the above derivations, we assume that $E-V(x)\neq0$.
For bound states and $|x|>a/2$ ($V_{11}(x)\equiv0$), the wave function $\psi_2(x)$  is exponentially decayed, i.e.,
\begin{align}\label{H0}
&\psi_2(x)=Ce^{-\lambda x} & x>a/2,\notag\\
&\psi_2(x)=De^{\lambda x} & x<-a/2,
\end{align}
where $\lambda=\sqrt{m^2-E^2}>0$.
For $|x|<a/2$, the wave function $\psi_2(x)$ can be
\begin{align}\label{18}
&\psi_2(x)=Ae^{ik x}+ Be^{-ik x} ,
\end{align}
where $k=\sqrt{(E-V)^2-m^2}>0$, $A$ and $B$ are two coefficients for oscillation parts of wave function.

In the following, we assume that the wave functions for all the three components have finite values (they are not infinitely large). Integrating both sides of Eq.(\ref{15}) near $x=\pm a/2$, we find
$\psi_2(x)$ and $\psi_1(x)+\psi_3(x)$ are continuous functions at $x=\pm a/2$ \cite{Fang2016,Chenxiaomei2019}. Furthermore, taking the relations of
\begin{align}\label{H0}
&\frac{-i\partial_x\psi_2(x)}{\sqrt{2}[E-m-V_{11}(x)]}=\psi_1(x),\notag\\
&\frac{-i\partial_x\psi_2(x)}{\sqrt{2}[E+m-V_{11}(x)]}=\psi_3(x)
\end{align}
into account,
we get  equations for coefficients A, B, C and D, i.e.,
\begin{align}\label{H0}
&Ae^{ika/2}+Be^{-ika/2}=Ce^{-\lambda a/2},\notag\\
&Ae^{-ika/2}+Be^{ika/2}=De^{-\lambda a/2},\notag\\
&\frac{2(E-V)(ikAe^{ika/2}-ikBe^{-ika/2})}{(E-m-V)(E+m-V)}=\frac{-2E\lambda Ce^{-\lambda a/2}}{E^2-m^2},\notag\\
&\frac{2(E-V)(ikAe^{-ika/2}-ikBe^{ika/2})}{(E-m-V)(E+m-V)}=\frac{2E\lambda De^{-\lambda a/2}}{E^2-m^2}.
\end{align}

Let the corresponding determinant vanish, we get an equation for bound state energy
\begin{align}\label{H0}
&[2E^4-4E^3V+2Em^2V-m^2V^2+2E^2(-m^2+V^2)]sin(ka)\notag\\
&+2E(V-E)\sqrt{(m^2-E^2)[(E-V)^2-m^2]}cos(ka)=0.
\end{align}
 Panel (b) of Fig.2 reports the bound state energy $E_B$ as a function of potential strength $V$. In all the figures, we set the width of potential well $a=1/m$.

When $|E-m|\ll m$, and $|V|\ll m$, the resulting bound state energy can be approximated by
\begin{align}\label{22}
E_B=E\simeq\mp m\pm\frac{ma^2V^2}{2}.
\end{align}
It is shown that near the thresholds of continuums $\pm m$, and in the case of  weak potential, the bound state energy (relative to the thresholds) is proportional to the square of strength ($E_B\pm m \propto V^2$), which is consistent with the ordinary one-dimensional bound state energies (see Fig.2).

 This is because for the square well potential of type I, the three diagonal matrix elements of  potential are equal (in the usual basis $|i=1,2,3\rangle$) [see Eq.(\ref{14})].  One can also view the potential energy in the energy band basis, i.e., $|-,k\rangle,|0,k\rangle$ and $|+,k\rangle$, consequently the diagonal form is also not changed approximately. Therefore for the three energy bands (flat band, upper and lower bands), they have three same potentials $V_{11}(x)$.


 For flat band, the states   are spatially localized.  Once the potential is turned on, the energy of the states within the range of potential well only shift a value of $V$ , i.e., $E=0\rightarrow E=V$ (see the black dashed line in panel (b) of Fig.2 ).  These resulting new states basically inherit the localized properties of the original flat band states. Therefore, we call them  trivial bound states (localized flat band states), which we are not interested in.

 In addition, near the thresholds of upper and lower continuums, the density of states has a similar divergence behaviors as that in ordinary one-dimensional cases, i.e., $1/\sqrt{|E-E_{th}|}\rightarrow\infty $ as  $E\rightarrow E_{th}$ \cite{Economou} [see Fig.1 and Eq.(\ref{7})]. So in such cases,  the bound state problem can be reduced into an ordinary one-dimensional bound state problem near the thresholds (especially for shallow bound states). Consequently the resulting bound state energies are very similar to the usual one-dimensional results, e.g., $E_B\pm m\propto V^2$.
 For a given potential strength $V$, the number of (non-trivial) bound states is always finite for the square well potential of type I (with three equal diagonal matrix elements).

\subsection{Square well potential of type II}
In this subsection, we assume the potential energy satisfies: $V_{11}(x)=V_{33}(x)\equiv0$ and
\begin{align}\label{H0}
&V_{22}(x)=0 & x>a/2,\notag\\
&V_{22}(x)=V & -a/2<x<a/2,\notag\\
&V_{22}(x)=0 & x<-a/2.
\end{align}

Now the Schr\"{o}dinger  equation can be rewritten as
\begin{align}\label{35}
&-i\partial_x\psi_2(x)/\sqrt{2}=[E-m]\psi_1(x),\notag\\
&-i\partial_x[\psi_1(x)+\psi_3(x)]/\sqrt{2}=[E-V_{22}(x)]\psi_2(x),\notag\\
&-i\partial_x\psi_2(x)/\sqrt{2}=[E+m]\psi_3(x).
\end{align}
Similarly as above subsections, we get an effective Schrodinger equation (second order differential equation)
\begin{align}\label{25}
\partial^{2}_x\psi_2(x)+\frac{[E-V_{22}(x)](E^2-m^2)}{E}\psi_2(x)=0.
\end{align}
In the above derivation, we assume that $E\neq0$.

\begin{figure}
\begin{center}
\includegraphics[width=1.0\columnwidth]{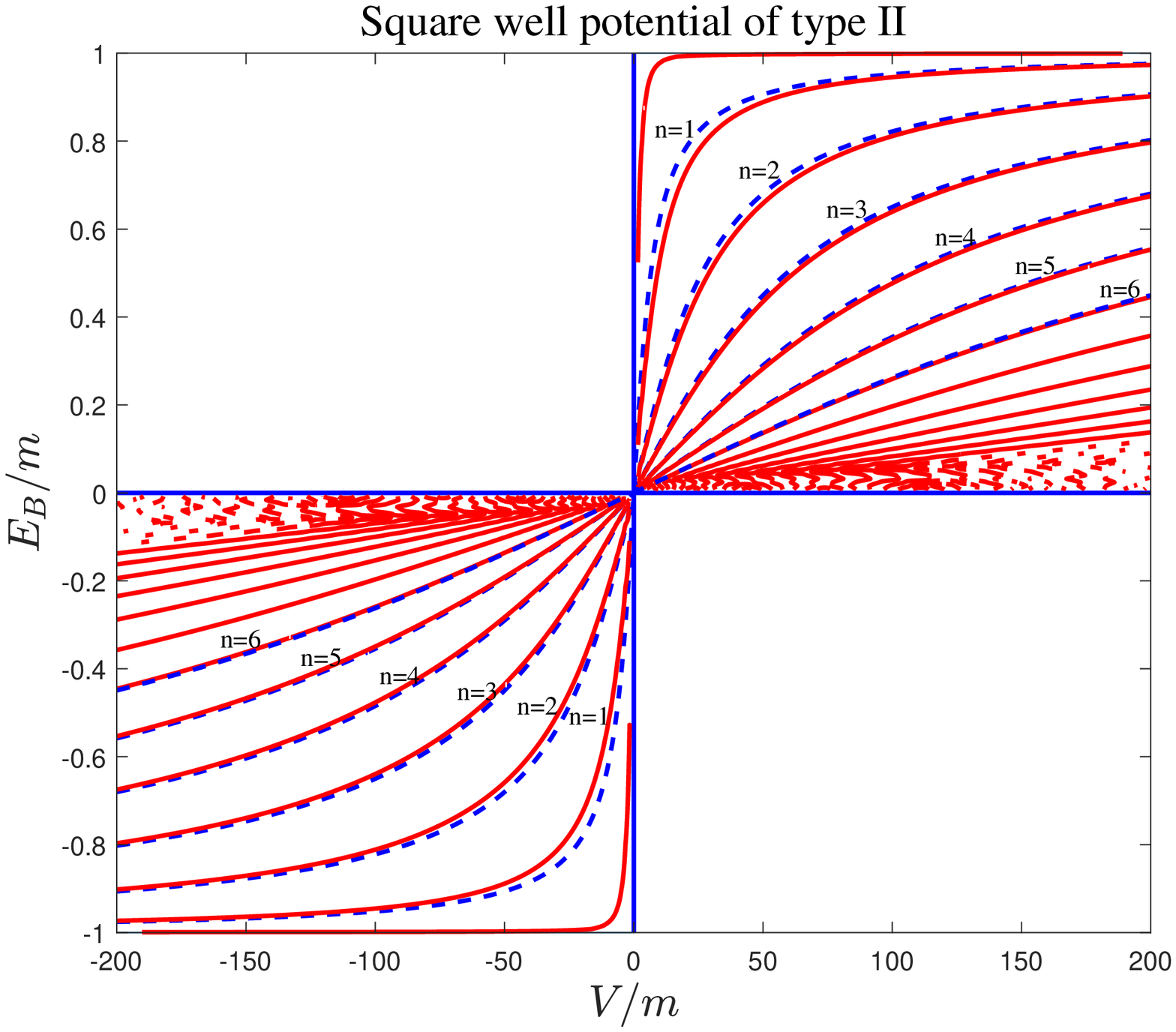}
\end{center}
\caption{ The bound state energy of square well potential of type II. The red solid lines are the the exact results of Eq.(\ref{42}). When $|E/V|\ll1$, the bound state energy can be approximated by Eq.(\ref{44}) (see the blue dashed lines).}
\label{schematic}
\end{figure}

\begin{figure}
\begin{center}
\includegraphics[width=1.0\columnwidth]{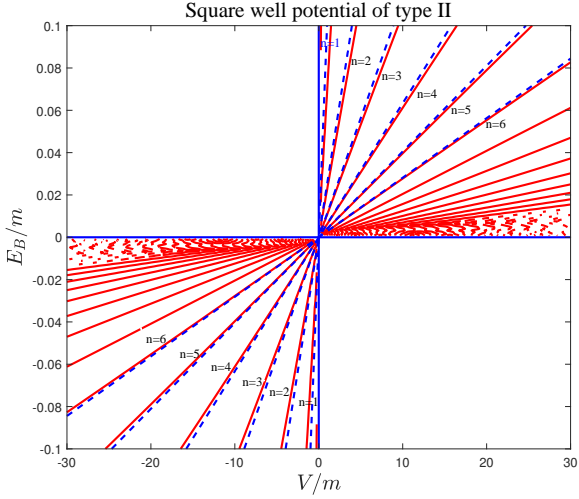}
\end{center}
\caption{ The bound state energy of square well potential of type II. The red solid lines are the the exact results of Eq.(\ref{42}). When $|E/m|\ll1$, the bound state energy is given by Eq.(\ref{46}) (see the blue dashed lines).} 
\label{schematic}
\end{figure}


For bound states and $|x|>a/2$, the wave function $\psi_2(x)$  can be written as
\begin{align}\label{26}
&\psi_2(x)=Ce^{-\lambda x} & x>a/2,\notag\\
&\psi_2(x)=De^{\lambda x} & x<-a/2,
\end{align}
where $\lambda=\sqrt{m^2-E^2}>0$.
For $|x|<a/2$, the wave function $\psi_2(x)$ is
\begin{align}\label{27}
&\psi_2(x)=Ae^{ik x}+ Be^{-ik x} ,
\end{align}
where $k=\sqrt{\frac{[E-V](E^2-m^2)}{E}}>0$.

Similarly integrating the both sides of Eq.(\ref{35}) near $x=\pm a/2$, we find
$\psi_2(x)$ and $\psi_1(x)+\psi_3(x)$ are continuous functions at $x=\pm a/2$. Furthermore, taking the relations
\begin{align}\label{H0}
&\frac{-i\partial_x\psi_2(x)}{\sqrt{2}(E-m)}=\psi_1(x),\notag\\
&\frac{-i\partial_x\psi_2(x)}{\sqrt{2}(E+m)}=\psi_3(x)
\end{align}
into account, it can be shown that the derivative $\partial_x\psi_2(x)$ at $x=\pm a/2$  is also continuous.

Using these continuity conditions of $\psi_2(x)$ at $x=\pm a/2$,
we get linear equations for $A$, $B$, $C$ and $D$
\begin{align}\label{H0}
&Ae^{ika/2}+Be^{-ika/2}=Ce^{-\lambda a/2},\notag\\
&Ae^{-ika/2}+Be^{ika/2}=De^{-\lambda a/2},\notag\\
&ikAe^{ika/2}-ikBe^{-ika/2}=-\lambda Ce^{-\lambda a/2},\notag\\
&ikAe^{-ika/2}-ikBe^{ika/2}=\lambda De^{-\lambda a/2}.
\end{align}

In order to have non-trivial solutions to the above linear equations, the corresponding determinant should vanish, i.e.,
\begin{align}
&Det\left|\begin{array}{cccc}
e^{ika/2} &e^{-ika/2}  & -e^{-\lambda a/2}& 0\\
e^{-ika/2}&e^{ika/2}& 0 &-e^{-\lambda a/2}\\
ike^{ika/2} &-ike^{-ika/2} & \lambda e^{-\lambda a/2}&0\\
ike^{-ika/2} &-ike^{ika/2} & 0&-\lambda e^{-\lambda a/2}
  \end{array}\right|=0.
\end{align}
Then, we get an equation of bound state energy
\begin{align}\label{42}
2E\sqrt{V/E-1}cos(ka)=(V-2E)sin(ka),
\end{align}
where $k=\sqrt{\frac{[E-V](E^2-m^2)}{E}}>0$. Taking $a=1/m$, the resulting bound state energies are reported in Figs.(3) and (4).

When $|E|\ll |V|$ , the above Eq.(\ref{42}) is simplified as
\begin{align}\label{32}
sin(\sqrt{\frac{(m^2-E^2)V}{E}}a)=0.
\end{align}
So the energy
\begin{align}\label{44}
E_n\equiv E_B=E=\frac{-n^2\pi^2+\sqrt{n^4\pi^4+4m^2a^4V^2}}{2a^2V},
\end{align}
where quantum number $n=1,2,3,...$ (see Fig.3). It is shown that there exist infinite bound states, which are generated from the flat band.  Due to infinitely large density of states of flat band, an arbitrarily small potential  $V$ indeed pulls some states out the the flat band and then they form bound states.
For a given $n$, when the potential is very strong ($V\rightarrow\pm\infty$), the bound state energy approaches the thresholds $\pm m$, which is very similar to the delta potential case. For weak potential, the energy is proportional to the potential strength, e.g., $E_n\propto V$.

 We should emphasize that the existence of infinite number of bound states is quite a universal phenomenon in the flat band system with a quite arbitrary potential well of type II (not limited to square well potential, for more explanations see \textbf{Appendix B}). As long as energy is small enough (very near the zero energy of flat band), the resulting effective potential ($\tilde{V}\propto V_{22}/E\rightarrow\infty$) would be arbitrarily strong, then the system would have infinite  bound states.

Furthermore when  $|E|\ll m$, the above Eq.(\ref{32}) becomes
\begin{align}\label{34}
sin(\sqrt{\frac{m^2V}{E}}a)=0,
\end{align}
so we get the energy
\begin{align}\label{46}
E_n=E_B=E\simeq\frac{m^2a^2V}{n^2\pi^2},
\end{align}
where quantum number $n=1,2,3,...$ (see Fig.4).
It is interesting that the bound state energy displays the hydrogen atom-like energy spectrum.
Outside the square potential well ($|x|>a/2$), the bound state wave functions  are all exponentially decayed [see Eq.(\ref{26})]. Within the potential well, i.e., $|x|<a/2$, the wave function would oscillate wildly  for large quantum $n$ [see Eq.(\ref{27})].
When $|E|\rightarrow0$, the density of states (DOS) can be approximately obtained
\begin{align}
DOS(E)=\frac{\Delta n}{\Delta E}\simeq |\frac{dn}{dE_n}|=\frac{ma|V|^{1/2}}{2\pi|E|^{3/2}}.
\end{align}
As $|E|\rightarrow0$, $DOS(E)\propto 1/|E|^{3/2}\rightarrow\infty$.

The origin of existence of the infinite number of bound states in square well potential is due to the $1/z$ singularity in the Green function and the $\delta(E)$  singularity of density of states induced by the flat band (see \textbf{Appendices  A and B}). For ordinary bound state problem in quantum mechanics textbook,  the energy $E$ usually appears in numerator of the square of wave vector, e.g., $k^2\sim E$.  While due to the $1/z$ of Green function in the flat band model, the  energy $E$ can appear in the denominator, e.g., $k^2\propto1/E$ [see Eqs.(\ref{34}) and (\ref{25})]. This is why an infinite number of bound states and even the hydrogen atom-like energy spectrum  for the square well potential of type II can appear. While for the square well potential of type I, the $1/z$  singularity does not appear in the  bound state equation [see Eqs.(\ref{16}) and (\ref{18})]. Consequently, the number of (non-trivial) bound states is always finite for a given potential strength $V$.
In order to see the importance of the $1/z$  singularity, we give more explanations in the \textbf{Appendices A and B}. For other types of square well potentials and general potential strengths, there also exist infinite bound states and $1/n^2$ energy spectrum [see \textbf{ Appendix C}].

\section{summary}
In conclusion, we  analytically calculate the free particle Green function and density of states for the one dimensional spin-1 Dirac model. It is found that the infinitely large density of states of flat band results in a $1/z$ singularity of Green function. Furthermore, we solve the bound state problems in the flat band system for several typical potential wells. It is found that the flat band and its ensuing $1/z$ singularity of Green function play crucial roles in the formations of bound states.
Due to $1/z$ singularity, the bound state energy shows a linearly dependence on the potential strength.    In addition, there exists an infinite number of (non-trivial) bound states for some kinds of square well potentials. Furthermore, the bound state energy displays hydrogen atom-like energy spectrum.
Our findings provide alternative ways to get the hydrogen atom-like energy spectrum in condensed matter physics.

Finally, in the presence of long-ranged Coulomb potential, there also exist infinite bound states \cite{Zhangyicai2,Zhangyicai3,Zhangyicai4}.
However, the bound state energy is usually proportional to $1/n$, which is different from $1/n^2$ of short-ranged square well potential.

The above findings  would provide some useful insights in the understanding of many-body physics of flat band. For example,
the infinite bound states induced by small potential imply that an arbitrarily small interaction
would dominate the physics that also reflects the fact that the flat band is not unstable under interactions.
Since the bound states can appear for both repulsive (positive strength) and attractive (negative strength) potentials, then one can expect that even
a repulsive interaction may result in superfluid/superconductor pairing states in flat band system \cite{Kobayashi}.

There are also other interesting questions, for example, how to introduce phase shift for the states in flat band, how the potential affects the phase shifts, what is the relationships between the phase shift and the number of bound states (Levinson theorem) \cite{Ma1985,Barton1985,Dong2000,Camblong2019,Calogeracos2004,Lin1999}, etc.,   need further investigations.

\appendix
\section{The relations between the  $1/z$  singularity of Green function and $\delta(E)$  singularity of density of states  }

There exists a general relation between (the trace of) Green function and density of states \cite{Economou}, i.e,
 \begin{align}\label{H0}
Tr[G_0(x,x,z)]=\int_{-\infty}^{\infty} dE \frac{\rho_0(x,E)}{z-E}.
\end{align}
In addition, if the density of states has a $\delta(E)$ singularity , e.g, $\rho_0(x,E)\propto\delta(E)$ , the the Green function would behave as
   \begin{align}\label{H0}
Tr[G_0(x,x,z)]=\int_{-\infty}^{\infty} dE \frac{\rho_0(x,E)}{z-E}\propto 1/z.
\end{align}

  Furthermore, for a general matrix element of Green function, if a flat band appears,  there also exists the $1/z$ singularity [see Eq.(\ref{6}) in the main text]. It is known that the Green function can be represented as (spectral decomposition)
\begin{align}\label{H0}
&G_{0,ij}(x,x',z)=\langle x;i|\frac{1}{z-H_0}|x';j\rangle\notag\\
&=\sum_{n,\lambda} \frac{\psi_{n,\lambda}(x;i)\psi^{*}_{n,\lambda}(x';j)}{z-E_{n,\lambda}}
\end{align}
where $n$ is energy band index, $\lambda$ denotes other quantum number for a given energy band $n$. Here eigenfunction $\psi_{n,\lambda}(x)$ satisfies $H_0\psi_{n,\lambda}(x)=E_{n,\lambda}\psi_{n,\lambda}(x)$.
If a flat band appears, for example, $E_{n,\lambda}\equiv 0$ for some specific index $n$, $G_{0,ij}(x,x',z)$
 includes a term $\sim \sum_{\lambda}\psi_{n,\lambda}(x;i)\psi^{*}_{n,\lambda}(x';j)/z$ , then it would display a $1/z$ singularity in general. It should be remarked that the above conclusions do not only apply to a free particle Hamiltonian $H_0$, but also apply to a total Hamiltonian $H=H_0+V_p$ [for example, see Eq.(\ref{hamiltonian})].

\section{The importance  of $1/z$ singularity of Green function }
In order to see the importance of $1/z$ singularity of Green function in bound state equations, we reformulate the bound state problem with Green function method  \cite{Economou}.

Fist of all, we introduce the full Green function $G(z)$
\begin{align}\label{H0}
G(z)=\frac{1}{z-H},
\end{align}
which corresponds to the total Hamiltonian $H=H_0+V_p$. In addition,
$G(z)$ satisfies the  Lippmann-Shwinger equation, i.e.,
\begin{align}\label{H0}
G(z)=G_0(z)+G_0(z)V_pG(z)
\end{align}
where
\begin{align}\label{H0}
G_0(z)=\frac{1}{z-H_0},
\end{align}
is the free particle Green function  in operator form.
Taking a potential energy in a form of
\begin{align}\label{V}
V(x)=V_{22}(x)\bigotimes|2\rangle\langle2|=\left[\begin{array}{ccc}
0 &0  & 0\\
0&V_{22}(x)& 0\\
0 &0 & 0
  \end{array}\right]
\end{align}
as an example (both the delta potential and square well potential of type II in Sec.\textbf{III} belong such a class),
the Lippmann-Shwinger equation (in coordinate representation) takes the following form
\begin{align}\label{H0}
&G_{ij}(x,x',z)=G_{0,ij}(x,x',z)\notag\\
&+\int dx_1 G_{0,i2}(x,x_1,z)V_{22}(x_1)G_{2j}(x_1,x',z).
\end{align}
Specifically, the matrix element
\begin{align}\label{H0}
&G_{22}(x,x',z)=G_{0,22}(x,x',z)\notag\\
&+\int dx_1 G_{0,22}(x,x_1,z)V_{22}(x_1)G_{22}(x_1,x',z).
\end{align}
The bound state solutions are determined by the corresponding homogenous integral equation, namely
 \begin{align}\label{H0}
G_{22}(x,x',z)=\int dx_1 G_{0,22}(x,x_1,z)V_{22}(x_1)G_{22}(x_1,x',z).
\end{align}


For simplifications,
taking a specific value $x'=0$, setting $z=E$ and  $G(x)\equiv G_{22}(x,0,E)$, the resulting integral equation is
\begin{align}\label{B8}
G(x)=\int dx_1 G_{0,22}(x,x_1,E)V_{22}(x_1)G(x_1).
\end{align}

Inserting $G_{0,22}(x,x',z)=\frac{\sqrt{m^2-z^2}}{2z}e^{-\sqrt{m^2-z^2}|x-x'|}$ [see Eq.(\ref{6}) in the main text] in  Eq.(\ref{B8}), we get
\begin{align}\label{B9}
G(x)=\frac{\sqrt{m^2-E^2}}{2E}\int dx_1 e^{-\sqrt{m^2-E^2}|x-x_1|}V_{22}(x_1)G(x_1).
\end{align}

It is  shown that $1/z$ singularity of Green function matrix element $G_{0,22}$  introduces a factor $1/E$, which results in the linear dependence on the potential strength in bound state energy .
For example, when the potential is a delta function, i.e.,  $V_{22}(x)=g\delta(x)$, then
the integral equation becomes
\begin{align}\label{H0}
G(x)=\frac{g\sqrt{m^2-E^2}}{2E}e^{-\sqrt{m^2-E^2}|x|}G(0).
\end{align}
Then
\begin{align}\label{H0}
G(0)=\frac{g\sqrt{m^2-E^2}}{2E}G(0),
\end{align}
and
\begin{align}\label{H0}
1=\frac{g\sqrt{m^2-E^2}}{2E}.
\end{align}
So we can get the bound state energy
\begin{align}\label{H0}
E_B=E=\frac{m g}{\sqrt{4+g^2}},
\end{align}
which is consistent with Eq.(\ref{13}) in the main text.

In the case of square well potential, the above Eq.(\ref{B9}) is essentially equivalent to effective Schr\"{o}dinger equation [Eq.(\ref{25}) in the main text].
This is because the differential equation Eq.(\ref{25})
\begin{align}
\partial^{2}_x\psi_2(x)+\frac{[E-V_{22}(x)](E^2-m^2)}{E}\psi_2(x)=0,
\end{align}
can be also transformed into an corresponding integral equation for bound state problem. Fist of all, let us write it into a form of an  ordinary second order Schr\"{o}dinger equation, e.g.,
\begin{align}\label{B15}
-\partial^{2}_x\psi_2(x)+\tilde{V}\psi_2(x)=\tilde{E}\psi_2(x),
\end{align}
where effective total energy $\tilde{E}$,  effective potential energy $\tilde{V}$
\begin{align}\label{B16}
&\tilde{E}=E^2-m^2, \notag\\
&\tilde{V}(x)=\tilde{E}-\frac{[E-V_{22}(x)](E^2-m^2)}{E}\notag\\
&=\frac{V_{22}(x)(E^2-m^2)}{E},
\end{align}
and $\tilde{E}<0$ for bound states. In the determining effective total energy, we assume potential energy $V_{22}(x)$ goes to zero as $x\rightarrow \pm \infty$.
 It shows that when $V/E>0$, the effective potential $\tilde{V}=V(E^2-m^2)/E<0$ is attractive, then there would exist bound states in the one dimension. Furthermore, if the energy is very small, e.g., $0<E/V\ll1$ and $|E|/m\ll1$, the effective potential strength can be arbitrarily large ($|\tilde{V}|\simeq|V|m/|E|\rightarrow \infty$), then there would exist infinite bound states.
  We see that factor $1/E$ plays crucial roles in the existence of infinite number of bound states.

It should be remarked that the existence of infinite  bound states is quite a universal phenomenon in the flat band system with a potential well of type II.
This is because as long as the energy $E$ is sufficient small,  the factor $1/E$ would result in an arbitrarily strong effective potential $\tilde{V}(x)$. Consequently, the strong effective potential would support  infinite bound states.

The integral equation corresponding to the above Eq.(\ref{B15}) is
\begin{align}\label{B17}
\psi_2(x)=\int dx_1 \tilde{G}_{0}(x,x_1,\tilde{E})\tilde{V}(x_1)\psi_2(x_1),
\end{align}
where the effective free particle Green function (corresponding to effective Hamiltonian $\tilde{H}_0=-\partial^{2}_{x}$ in one dimension) \cite{Economou}
 \begin{align}
&\tilde{G}_{0}(x,x_1,\tilde{E})=\frac{-e^{-\sqrt{-\tilde{E}}|x-x_1|}}{2\sqrt{-\tilde{E}}}\notag\\
&=\frac{-e^{-\sqrt{m^2-E^2}|x-x_1|}}{2\sqrt{m^2-E^2}}.
\end{align}

Inserting it in Eq.(\ref{B17}), then
\begin{align}\label{B19}
\psi_2(x)=\frac{\sqrt{m^2-E^2}}{2E}\int dx_1 e^{-\sqrt{m^2-E^2}|x-x_1|}V_{22}(x_1)\psi_2(x_1).
\end{align}
It is found that $G(x)$ and $\psi_2(x)$ satisfy the same integral equation for bound states [see Eqs.(\ref{B9}) and (\ref{B19})].
During the above derivations, we see that $1/z$ singularity in Green function is very important in the bound state equation. The $1/z$ singularity have close relations to the  factor $E$ in the denominator of the effective potential $\tilde{V}$ [see Eqs.(\ref{B16}), (\ref{B9}) and (\ref{B19}) ].
  Most of  the peculiar behaviors of bound state energies, for example linear dependence on potential strength, infinite number of bound states and hydrogen atom-like spectrum can be attributed to the $1/z$ singularity of Green function.

\begin{figure}
\begin{center}
\includegraphics[width=1.0\columnwidth]{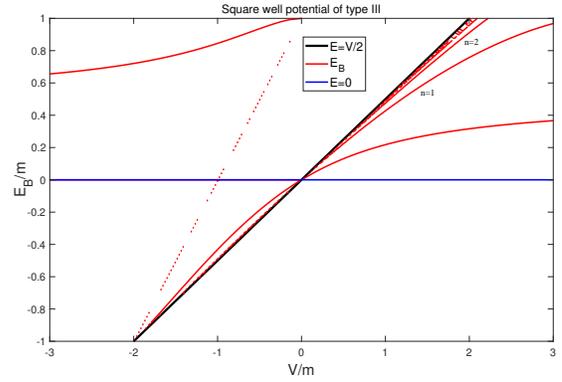}
\end{center}
\caption{ The bound state energy of square well potential of type III. The red solid lines are the the exact results of Eq.(\ref{C9}). When $V<0$ and $|E/m|\ll1$, the  bound state energy can be approximated by the Eq.(\ref{C10}). Near the black line ($E=V/2$), there also exist infinite bound states.}
\end{figure}

\begin{figure}
\begin{center}
\includegraphics[width=1.0\columnwidth]{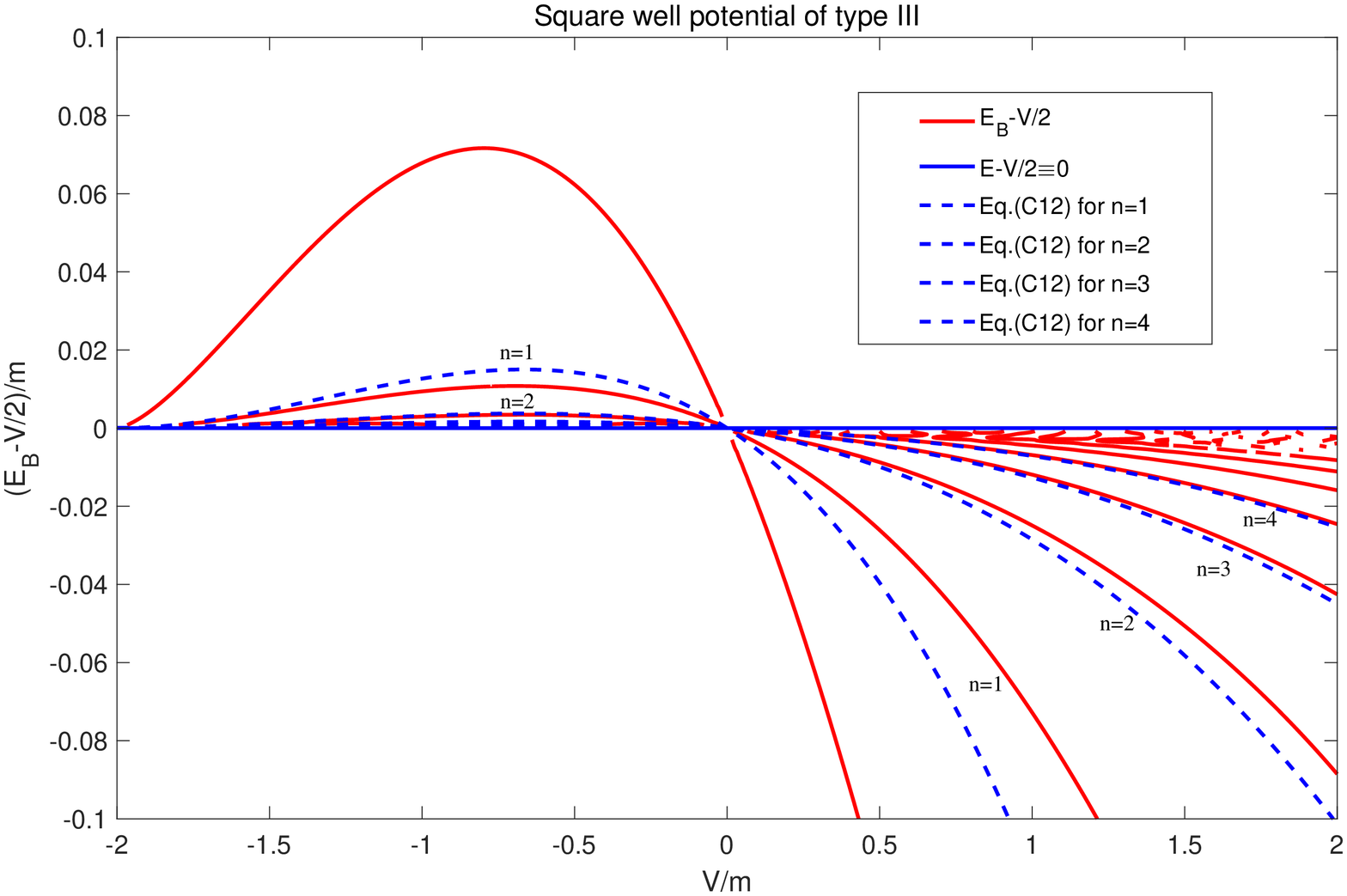}
\end{center}
\caption{ The bound state energy (relative to $V/2$) of square well potential of type III. The red solid lines are the the exact results of Eq.(\ref{C9}). Further more, when $|E-V/2|/m\ll1$, the bound state energy is approximated by Eq.(\ref{C12}) (see the blue dashed lines).}
\end{figure}

\section{Square well potential of type III}
In this appendix, we discuss the bound states for other types of square well potentials. First of all, we assume all the matrix elements $V_{11}(x)$, $V_{22}(x)$  and $V_{22}(x)$ are square well potentials with same width $a$, i.e.,
\begin{align}\label{H0}
&V_{11}(x)=0 & x>a/2,\notag\\
&V_{11}(x)=V_{11} & -a/2<x<a/2,\notag\\
&V_{11}(x)=0 & x<-a/2,\notag\\
&V_{22}(x)=0 & x>a/2,\notag\\
&V_{22}(x)=V_{22} & -a/2<x<a/2,\notag\\
&V_{22}(x)=0 & x<-a/2,\notag\\
&V_{33}(x)=0 & x>a/2,\notag\\
&V_{33}(x)=V_{33} & -a/2<x<a/2,\notag\\
&V_{33}(x)=0 & x<-a/2,
\end{align}
where $V_{11}$, $V_{22}$ and $V_{33}$ are potential strengths of  $V_{11}(x)$, $V_{22}(x)$ and $V_{22}(x)$, respectively.

Now the Schr\"{o}dinger  equation can be rewritten as
\begin{align}\label{24}
&-i\partial_x\psi_2(x)/\sqrt{2}=[E-m-V_{11}(x)]\psi_1(x),\notag\\
&-i\partial_x[\psi_1(x)+\psi_3(x)]/\sqrt{2}=[E-V_{22}(x)]\psi_2(x),\notag\\
&-i\partial_x\psi_2(x)/\sqrt{2}=[E+m-V_{33}(x)]\psi_3(x).
\end{align}
Similarly eliminating $\psi_1(x)$ and $\psi_3(x)$, we get
\begin{align}\label{H0}
& \partial^{2}_x\psi_2(x)+\partial_x\psi_2(x)\partial_x [ln(\frac{2E-[V_{11}(x)+V_{33}(x)]}{[E-m-V_{11}(x)][E+m-V_{33}(x)]})]\notag\\
&+\frac{2(E-V_{22}(x))(E+m-V_{33})[E-m-V_{11}(x)]}{2E-[V_{11}(x)+V_{33}(x)]}\psi_2(x)=0.
\end{align}

When $|x|>a/2$, the bound state wave function $\psi_2(x)$  can be written as
\begin{align}\label{H0}
&\psi_2(x)=Ce^{-\lambda x} & x>a/2,\notag\\
&\psi_2(x)=De^{\lambda x} & x<-a/2,
\end{align}
where $\lambda=\sqrt{m^2-E^2}>0$.
For $|x|<a/2$, the wave function $\psi_2(x)$ is
\begin{align}\label{H0}
&\psi_2(x)=Ae^{ik x}+ Be^{-ik x} ,
\end{align}
where
\begin{align}\label{C6}
k=\sqrt{\frac{2(E-V_{22})(E-m-V_{11})[E-m-V_{33}]}{2E-(V_{11}+V_{33})}}>0.
\end{align}

Integrating the both sides of Eq.(\ref{24}) near $x=\pm a/2$, we find
$\psi_2(x)$ and $\psi_1(x)+\psi_3(x)$ are continuous functions near $x=\pm a/2$. Further more, taking the relations
\begin{align}\label{H0}
&\frac{-i\partial_x\psi_2(x)}{\sqrt{2}[E-m-V_{11}(x)]}=\psi_1(x),\notag\\
&\frac{-i\partial_x\psi_2(x)}{\sqrt{2}[E+m-V_{33}(x)]}=\psi_3(x)
\end{align}
into account,
we get
\begin{align}\label{H0}
&Ae^{ika/2}+Be^{-ika/2}=Ce^{-\lambda a/2},\notag\\
&Ae^{-ika/2}+Be^{ika/2}=De^{-\lambda a/2},\notag\\
&\frac{[2E-(V_{11}+V_{33})](ikAe^{ika/2}-ikBe^{-ika/2})}{(E-m-V_{11})(E+m-V_{33})}=\frac{-2E\lambda Ce^{-\lambda a/2}}{E^2-m^2},\notag\\
&\frac{[2E-(V_{11}+V_{33})](ikAe^{-ika/2}-ikBe^{ika/2})}{(E-m-V_{11})(E+m-V_{33})}=\frac{2E\lambda De^{-\lambda a/2}}{E^2-m^2}.
\end{align}

Let the corresponding determinant vanish, we can get
bound state energies. When $V_{11}=V_{22}=V_{33}=V$, the bound state energy would be reduced to Eq.(21) of potential of type I.
 While when  $V_{11}=V_{33}\equiv0, V_{22}=V$, the bound state energy is given by Eq.(31) of type II.
We see for a given potential strength $V$, there exist finite bound states in case of type I, while the system has infinite bound states in the case of type II.

From the discussions of  \textbf{Appendix B}, we know that  a crucial requirement for existence of infinite bound states is that the effective potential [or  wave vector $k$ in Eq.(\ref{C6})] can become infinitely large. So if the denominator of $k^2$ [see Eq.(\ref{C6})] can not be canceled by the numerator, the wave vector $k$ and  effective potential can be arbitrarily large for some $E$. Then, there would exist infinite bound states. While when the denominator of $k^2$  can be canceled  by the its numerator, the wave vector $k$ is always finite. In such a case, the system only have finite bound states, e.g., the case of $V_{11}=V_{22}=V_{33}=V$ (the case of type I).

In the following, we investigate the case of $V_{22}=V_{33}\equiv0,V_{11}=V$, where  the denominator of $k^2$ can not be canceled by its numerator. Then it is expected that there would exist infinite bound states.
In the whole manuscript, we would refer such a kind of potential as potential of type III.
The bound state energy equation is
\begin{align}\label{C9}
&2(-2E+V)\sqrt{\frac{m-E}{m+E}}kcos(ka)\notag\\
&+(4E^2-4Em-3EV+mV)sin(ka)=0,
\end{align}
where $k=\sqrt{\frac{2E(E+m)[E-m-V]}{2E-V}}>0$.
The results are reported in Figs.5 and 6.
When $V<0$ and  $E\rightarrow E_{th}=m$, the resulting bound state energy
\begin{align}\label{C10}
E_B\simeq m-\frac{ma^2V^2}{2},
\end{align}
which is also consistent with result of one-dimensional bound states [see Figs.2, 5 and Eq.(\ref{22})].

In addition, when energy $E$ approaches $V/2$, e.g., $2E-V\rightarrow0$, the wave vector $k\rightarrow \infty$. Consequently, there exist infinite bound states near the line $E=V/2$ (see black line of Fig.5).
When $|E-V/2|/m\ll1$ , the above Eq.(\ref{C9}) is simplified into
\begin{align}\label{H0}
&sin(\sqrt{\frac{2E(E+m)[E-m-V]}{2E-V}}a)\notag\\
&\simeq sin(\sqrt{\frac{-V(m+V/2)^2}{2E-V}}a)=0.
\end{align}
Then we get the energy
\begin{align}\label{C12}
E_B=E\simeq\frac{V}{2}-\frac{V(m+V/2)^2a^2}{2n^2\pi^2},
\end{align}
where $n=1,2,3,...$
There also exist infinite bound states for arbitrarily small potential strength $V$ (see Fig.5). 
 The bound state energy (relative to $V/2$) also displays the hydrogen atom-like energy spectrum (see Fig.6).

The above discussions indicate that, for other types of square well potentials and general potential strengths, there also exist infinite bound states and $1/n^2$ energy spectrum 


%


\section*{Acknowledgements}
YC Zhang thank Prof. Shizhong Zhang for useful discussions.
This work was supported by the NSFC under Grants Nos.
11874127, 11504095. YC. Zhang also acknowledges the supports of startup grant from Guangzhou University.



\begin{thebibliography}{0}%
\makeatletter
\providecommand \@ifxundefined [1]{%
 \@ifx{#1\undefined}
}%
\providecommand \@ifnum [1]{%
 \ifnum #1\expandafter \@firstoftwo
 \else \expandafter \@secondoftwo
 \fi
}%
\providecommand \@ifx [1]{%
 \ifx #1\expandafter \@firstoftwo
 \else \expandafter \@secondoftwo
 \fi
}%
\providecommand \natexlab [1]{#1}%
\providecommand \enquote  [1]{``#1''}%
\providecommand \bibnamefont  [1]{#1}%
\providecommand \bibfnamefont [1]{#1}%
\providecommand \citenamefont [1]{#1}%
\providecommand \href@noop [0]{\@secondoftwo}%
\providecommand \href [0]{\begingroup \@sanitize@url \@href}%
\providecommand \@href[1]{\@@startlink{#1}\@@href}%
\providecommand \@@href[1]{\endgroup#1\@@endlink}%
\providecommand \@sanitize@url [0]{\catcode `\\12\catcode `\$12\catcode
  `\&12\catcode `\#12\catcode `\^12\catcode `\_12\catcode `\%12\relax}%
\providecommand \@@startlink[1]{}%
\providecommand \@@endlink[0]{}%
\providecommand \url  [0]{\begingroup\@sanitize@url \@url }%
\providecommand \@url [1]{\endgroup\@href {#1}{\urlprefix }}%
\providecommand \urlprefix  [0]{URL }%
\providecommand \Eprint [0]{\href }%
\providecommand \doibase [0]{http://dx.doi.org/}%
\providecommand \selectlanguage [0]{\@gobble}%
\providecommand \bibinfo  [0]{\@secondoftwo}%
\providecommand \bibfield  [0]{\@secondoftwo}%
\providecommand \translation [1]{[#1]}%
\providecommand \BibitemOpen [0]{}%
\providecommand \bibitemStop [0]{}%
\providecommand \bibitemNoStop [0]{.\EOS\space}%
\providecommand \EOS [0]{\spacefactor3000\relax}%
\providecommand \BibitemShut  [1]{\csname bibitem#1\endcsname}%
\let\auto@bib@innerbib\@empty
\end{thebibliography}%


\begin{thebibliography}{10}

\bibitem{Sutherland1986} Bill Sutherland, Localization of electronic wave functions due to local topology, Phys. Rev. B \textbf{34}, 5208 (1986).
\bibitem{Vidal1998} Julien Vidal, R\'{e}my Mosseri, and Benoit Doucot, Aharonov-Bohm Cages in Two-Dimensional Structures, Phys. Rev. Lett. \textbf{81}, 5888 (1998).

\bibitem{Bergman2008}Doron L. Bergman, Congjun Wu, and Leon Balents, Band touching from real-space topology in frustrated hopping models, Phys. Rev. B \textbf{78}, 125104 (2008).
\bibitem{Bercioux2009}
D. Bercioux, D. F. Urban, H. Grabert, and W. H\"{a}usler, Massless Dirac-Weyl fermions in a $T_3$
 optical lattice, Phys. Rev. A \textbf{80}, 063603 (2009).
\bibitem{Leykam2018} Daniel Leykam, Alexei Andreanov and Sergej Flach, Artificial flat
band systems: from lattice models to experiments, Advances in Physics: X, \textbf{3:1}, 1473052  (2018).
\bibitem{Moessner2011} Bal\'{a}zs D\'{o}ra, Janik Kailasvuori, and R. Moessner, Lattice generalization of the Dirac equation to general spin and the role of the flat band, Phys. Rev. B \textbf{84}, 195422 (2011).
\bibitem{Raoux2014} A. Raoux, M. Morigi, J.-N. Fuchs, F. Pi\'{e}chon, and G. Montambaux,
From Dia to Paramagnetic Orbital Susceptibility of Massless Fermions,
Phys. Rev. Lett. \textbf{112}, 026402 (2014).
\bibitem{Peotta2015} S. Peotta,  and  P. T\"{o}rm\"{a},  Superfluidity in topologically nontrivial flat bands. Nat.Commun. \textbf{6}, 8944 (2015).
\bibitem{Cao2018} Yuan Cao, et.al., Unconventional superconductivity in magic-angle graphene superlattices, Nature \textbf{556}, 43 (2018).

\bibitem{Hazra2019} Tamaghna Hazra, Nishchhal Verma,and Mohit Randeria, Bounds on the Superconducting Transition Temperature: Applications to Twisted Bilayer Graphene and Cold Atoms, Phys. Rev. X \textbf{9}, 031049 (2019).
  \bibitem{Xu2020} Hong-Ya Xu and Ying-Cheng Lai, Anomalous chiral edge states in spin-1 Dirac quantum dots, Phys. Rev. Research \textbf{2}, 013062 (2020).
\bibitem{Wang2011} Fa Wang and Ying Ran, Nearly flat band with Chern number
C=2 on the dice lattice,
  Phys. Rev. B \textbf{84}, 241103(R) (2011).

 \bibitem{Illes2016} E. Illes and E. J. Nicol, Magnetic properties of the $\alpha-T_3$
 model: Magneto-optical conductivity and the Hofstadter butterfly, Phys. Rev. B \textbf{94}, 125435 (2016).

\bibitem{Chen2019} Yan-Ru Chen, Yong Xu, Jun Wang, Jun-Feng Liu, and Zhongshui Ma, Enhanced magneto-optical response due to the flat band in nanoribbons made from the
$\alpha-T_3$ lattice
Phys. Rev. B \textbf{99}, 045420(2019).

 \bibitem{Wuyurong2021} Yu-Rong Wu and Yi-Cai Zhang, Superfluid states in $\alpha-T_3$ lattice, Chinese Phys. B \textbf{30}, 060306 (2021).


\bibitem{Mielke1999} Andreas Mielke, Ferromagnetism in Single-Band Hubbard Models with a Partially Flat Band, Phys. Rev. Lett. \textbf{82}, 4312(1999).







\bibitem{Leykam2017} Daniel Leykam, Joshua D. Bodyfelt, Anton S. Desyatnikov and Sergej Flach, Localization of weakly disordered flat band states. Eur. Phys. J. B \textbf{90}, 1 (2017).




\bibitem{Shen2010} R. Shen, L. B. Shao, Baigeng Wang, and D. Y. Xing, Single Dirac cone with a flat band touching on line-centered-square optical lattices, Phys. Rev. B \textbf{81}, 041410 (2010).

\bibitem{Urban2011} Daniel F. Urban, Dario Bercioux, Michael Wimmer, Wolfgang H\"{a}usler, Barrier transmission of Dirac-like pseudospin-one particles, Phys. Rev. B \textbf{84}, 115136 (2011).

\bibitem{Illes2017} E. Illes and E. J. Nicol, Klein tunneling in the
$\alpha-T_3$ model, Phys. Rev. B \textbf{95}, 235432 (2017).

\bibitem{Fang2016} A. Fang, Z. Q. Zhang, Steven G. Louie, and C. T. Chan, Klein tunneling and supercollimation of pseudospin-1 electromagnetic waves, Phys. Rev. B \textbf{93}, 035422 (2016).


\bibitem{Ocampo2017} Y. Betancur-Ocampo, G. Cordourier-Maruri, V. Gupta, and R. de Coss, Super-Klein tunneling of massive pseudospin-one particles, Phys. Rev. B \textbf{96}, 024304 (2017).
\bibitem{BERGHOLTZ} E. J. Bergholtz, Zhao Liu, Topological flat bnad models and fractional Chern insulator, International Journal of Modern Physics B
\textbf{27}, 1330017 (2013).

\bibitem{Yang2012} Shuo Yang, Zheng-Cheng Gu, Kai Sun, and S. Das Sarma, Topological flat band models with arbitrary Chern numbers,
Phys. Rev. B \textbf{86}, 241112(R) (2012).
\bibitem{Ghosh} Tutul Biswas and Tarun Kanti Ghosh, Dynamics of a quasiparticle in the $\alpha-T_3$
model: role of pseudospin polarization and
transverse magnetic field on zitterbewegung, J. Phys.: Condens. Matter \textbf{30}, 075301 (2018).

 \bibitem{Tovmasyan2018} Murad Tovmasyan, Sebastiano Peotta, Long Liang, P\"{a}ivi T\"{o}rm\"{a}, and Sebastian D. Huber, Preformed pairs in flat Bloch bands
Phys. Rev. B \textbf{98}, 134513 (2018).

\bibitem{Volovik2019} Volovik, G.E. Flat Band and Planckian Metal. Jetp Lett. \textbf{110}, 352-353 (2019).
\bibitem{Xie2020} Fang Xie, Zhida Song, Biao Lian, and B. Andrei Bernevig,  Topology-Bounded Superfluid Weight in Twisted Bilayer Graphene, Phys. Rev. Lett. \textbf{124}, 167002 (2020).

\bibitem{Julku2020} A. Julku, T. J. Peltonen, L. Liang, T. T. Heikkil\"{a}, and P. T\"{o}rm\"{a}
, Superfluid weight and Berezinskii-Kosterlitz-Thouless transition temperature of twisted bilayer graphene, Phys. Rev. B  \textbf{101}, 060505(R) (2020).
\bibitem{Hu2019} Xiang Hu, Timo Hyart, Dmitry I. Pikulin, and Enrico Rossi,  Topology-Bounded Superfluid Weight in Twisted Bilayer Graphene, Phys. Rev. Lett. \textbf{123}, 237002 (2019).
\bibitem{Kopnin2011} N. B. Kopnin, T. T. Heikkila, and G. E. Volovik
, High-temperature surface superconductivity in topological flat-band systems, Phys. Rev. B \textbf{83}, 220503(R) (2011).
\bibitem{Mukherjee} Sebabrata Mukherjee, et al., Observation of a Localized Flat-Band State in a Photonic Lieb Lattice, Phys. Rev. Lett. \textbf{114}, 245504 (2015).


\bibitem{Wu2008} Congjun Wu and S. Das Sarma, $p_{x,y}$-orbital counterpart of graphene: Cold atoms in the honeycomb optical lattice, Phys. Rev. B  \textbf{77}, 235107 (2008).
\bibitem{Zhang2010} Shizhong Zhang, Hsiang-hsuan Hung, and Congjun Wu, Proposed realization of itinerant ferromagnetism in optical lattices, Phys. Rev.  A \textbf{82}, 053618 (2010).
\bibitem{Iglovikov2014} V. I. Iglovikov, et.al., Superconducting transitions in flat-band systems, Phys. Rev. B \textbf{90}, 094506 (2014).

\bibitem{Julku2016} Aleksi Julku, et. al.,  Geometric Origin of Superfluidity in the Lieb-Lattice Flat Band, Phys. Rev. Lett. \textbf{117}, 045303 (2016).

 \bibitem{Liang2017} Long Liang, et.al.,  Band geometry, Berry curvature, and superfluid weight, Phys. Rev. B \textbf{95}, 024515 (2017).

  \bibitem{Iskin2019} Iskin, M. Origin of fat-band superfuidity on the Mielke checkerboard lattice. Phys. Rev. A \textbf{99}, 053608 (2019).

\bibitem{Wu2021} Yu-Rong Wu, Xiao-Fei Zhang, Chao-Fei Liu, Wu-Ming Liu and Yi-Cai Zhang, Superfluid density and collective modes of fermion superfluid in dice lattice, Sci Rep \textbf{11}, 13572 (2021)


\bibitem{Economou} E. N. Economou, Green's Functions in Quantum Physics,(Springer-Verlag Berlin Heidelberg, Third Edition, 2006).
\bibitem{Landau} L D Landau, E M Lifshitz, Quantum mechanics: non-relativistic theory, (Pergamon Press, Third Revised Edition, 1977).



\bibitem{Gorbar2019} E. V. Gorbar, V. P. Gusynin, and D. O. Oriekhov, Electron states for gapped pseudospin-1 fermions in the field of a charged impurity, Phys. Rev. B \textbf{99} 155124 (2019).

\bibitem{Pottelberge2020} R. Van Pottelberge, Comment on ``Electron states for gapped pseudospin-1 fermions in the field of a charged impurity", Phys. Rev. B \textbf{101}, 197102 (2020).


\bibitem{Zhang2013} Yi-Cai Zhang, Shu-Wei Song, Chao-Fei Liu, and Wu-Ming Liu, Zitterbewegung effect in spin-orbit-coupled spin-1 ultracold atoms, Phys. Rev. A \textbf{87}, 023612 (2013).
\bibitem{Chenxiaomei2019} Xiaomei Chen and Rui Zhu, Quantum Pumping with Adiabatically Modulated
Barriers in Three-Band Pseudospin-1
Dirac-Weyl Systems, Entropy, \textbf{21}, 209 (2019).



\bibitem{Huang2011} X Huang, Y Lai, ZH Hang, H Zheng, CT Chan, Dirac cones induced by accidental degeneracy in photonic crystals and zero-refractive-index materials, Nature Mater \textbf{ 10}, 582¨C586 (2011)

\bibitem{Chan2012} C. T. Chan, Zhi Hong Hang, and Xueqin Huang, Dirac Dispersion in Two-Dimensional Photonic Crystals, Adv. Optoelectron. 2\textbf{012}, 313984 (2012)



\bibitem{Zhang2014}  YC. Zhang, SW. Song,  and  WM. Liu, The confinement induced resonance in spin-orbit coupled cold atoms with Raman coupling. Sci Rep \textbf{4}, 4992 (2014).




\bibitem{Ma1985} Zhong-Qi Ma and Guang-Jiong Ni, Levinson theorem for Dirac particles, Phys. Rev. D \textbf{31}, 1482 (1985).


\bibitem{Barton1985} G. Barton, Levinson's theorem in one dimension: heuristics, J. Phys. A: Math. Gen. \textbf{18}, 479 (1985).
\bibitem{Lin1999}  Q. G. Lin, Levinson theorem for Dirac particles in one dimension. Eur. Phys. J. D \textbf{7}, 515 (1999).

\bibitem{Dong2000}  SH. Dong, ZQ.  Ma, Levinson's Theorem for the Schr\"{o}dinger Equation in One Dimension. International Journal of Theoretical Physics \textbf{39}, 469 (2000).

\bibitem{Camblong2019} H. E. Camblong, et.al., Spectral density, Levinson's theorem, and the extra term in the second virial coefficient
for the one-dimensional¦Ä-function potential, Phys. Rev.  A \textbf{100}, 062110 (2019).


\bibitem{Calogeracos2004} Alex Calogeracos, Norman Dombey, Strong Levinson Theorem for the Dirac Equation, Phys. Rev. Lett. \textbf{93}, 180405 (2004).

\bibitem{Zhangyicai2} Yi-Cai Zhang, Wave function collapses and 1/n energy spectrum induced by a Coulomb potential in a one-dimensional flat band system, \url{https://iopscience.iop.org/article/10.1088/1674-1056/ac3653} (2021).
\bibitem{Zhangyicai3} Yi-Cai Zhang, Infinite bound states and 1/n energy spectrum induced by a Coulomb potential of
type III in a flat band system, 2022, Phys. Scr. \textbf{97} 015401. \url{https://dx.doi.org/10.1088/1402-4896/ac46f4}.

\bibitem{Zhangyicai4} Yi-Cai Zhang, Bound States in the Continuum (BIC) Protected By Self-Sustained Potential Barriers in a Flat Band System, \url{https://www.researchgate.net/publication/356223517}(2021).

\bibitem{Kobayashi} Kobayashi K, Okumura M, Yamada S, Machida M and Aoki H 2016 Superconductivity in repulsively interacting fermions on a diamond chain: Flat-band-induced pairing Phys. Rev. B \textbf{94} 214501





%
%
%
%
%
%


























%

%
%
%
%
%
%





%
%
%
%
%
%
%
%
%
%
%
%
%
%
%
%
%
%
%
%
%
%

%
%
%
%

%
%
%
%
%







%



















%









%
%
%
%
%
%
%
%
%
%









%
%
%



%
%
%



































\end{thebibliography}
\end{document}